\documentclass[twocolumn]{aastex63}

\usepackage{amssymb,amsmath,amsthm}
\usepackage{graphicx}
\usepackage{mathrsfs}
\usepackage{booktabs}
\usepackage{multirow}
\usepackage{empheq}
\usepackage{verbatim} 
\usepackage{mathtools}
\usepackage{hyperref}
\usepackage{xcolor}
\usepackage{xspace}

\newcommand{\kms}{km\,s$^{-1}$\xspace}  
\newcommand{\kmsyr}{km\,s$^{-1}$\,yr$^{-1}$\xspace}  
\newcommand{\kmsmpc}{km\,s$^{-1}$\,Mpc$^{-1}$\xspace}   

\shorttitle{Megamaser Cosmology Project XIII}
\shortauthors{Pesce et al.}

\begin{document}

\title{The Megamaser Cosmology Project. XIII. Combined Hubble constant constraints}

\correspondingauthor{Dominic~W.~Pesce}
\email{dpesce@cfa.harvard.edu}

\author[0000-0002-5278-9221]{D.~W.~Pesce}
\affiliation{Center for Astrophysics $|$ Harvard \& Smithsonian, 60 Garden Street, Cambridge, MA 02138, USA}
\affiliation{Black Hole Initiative at Harvard University, 20 Garden Street, Cambridge, MA 02138, USA}

\author{J.~A.~Braatz}
\affiliation{National Radio Astronomy Observatory, 520 Edgemont Road, Charlottesville, VA 22903, USA}

\author{M. J. Reid}
\affiliation{Center for Astrophysics $|$ Harvard \& Smithsonian, 60 Garden Street, Cambridge, MA 02138, USA}

\author{A. G. Riess}
\affiliation{Department of Physics and Astronomy, Johns Hopkins University, Baltimore, MD, USA}
\affiliation{Space Telescope Science Institute, Baltimore, MD, USA}

\author{D. Scolnic}
\affiliation{Department of Physics, Duke University, 120 Science Drive, Durham, NC, 27708, USA}

\author{J. J. Condon}
\affiliation{National Radio Astronomy Observatory, 520 Edgemont Road, Charlottesville, VA 22903, USA}

\author{F. Gao}
\affiliation{Key Laboratory for Research in Galaxies and Cosmology, Shanghai Astronomical Observatory, Chinese Academy of Science, Shanghai 200030, China}
\affiliation{National Radio Astronomy Observatory, 520 Edgemont Road, Charlottesville, VA 22903, USA}

\author{C. Henkel}
\affiliation{Max-Planck-Institut f\"ur Radioastronomie, Auf dem H\"ugel 69, D-53121 Bonn, Germany}
\affiliation{Astronomy Department, Faculty of Science, King Abdulaziz University, P.O. Box 80203, Jeddah 21589, Saudi Arabia}

\author{C. M. V. Impellizzeri}
\affiliation{Joint ALMA Observatory, Alonso de Cordova 3107, Vitacura, Santiago, Chile}
\affiliation{National Radio Astronomy Observatory, 520 Edgemont Road, Charlottesville, VA 22903, USA}

\author{C. Y. Kuo}
\affiliation{Department of Physics, National Sun Yat-Sen University, No.70, Lianhai Rd., Gushan Dist., Kaohsiung City 804, Taiwan (R.O.C.)}
\affiliation{Academia Sinica Institute of Astronomy and Astrophysics, P.O. Box 23-141, Taipei 10617, Taiwan (R.O.C.)}

\author{K. Y. Lo}\altaffiliation{Deceased}
\affiliation{National Radio Astronomy Observatory, 520 Edgemont Road, Charlottesville, VA 22903, USA}

\begin{abstract}
We present a measurement of the Hubble constant made using geometric distance measurements to megamaser-hosting galaxies. We have applied an improved approach for fitting maser data and obtained better distance estimates for four galaxies previously published by the Megamaser Cosmology Project: UGC 3789, NGC 6264, NGC 6323, and NGC 5765b.  Combining these updated distance measurements with those for the maser galaxies CGCG 074-064 and NGC 4258, and assuming a fixed velocity uncertainty of 250\,\kms associated with peculiar motions, we constrain the Hubble constant to be $H_0 = 73.9 \pm 3.0$\,\kmsmpc independent of distance ladders and the cosmic microwave background.  This best value relies solely on maser-based distance and velocity measurements, and it does not use any peculiar velocity corrections.  Different approaches for correcting peculiar velocities do not modify $H_0$ by more than ${\pm}1{\sigma}$, with the full range of best-fit Hubble constant values spanning 71.8--76.9\,\kmsmpc.  We corroborate prior indications that the local value of $H_0$ exceeds the early-Universe value, with a confidence level varying from 95--99\% for different treatments of the peculiar velocities.
\end{abstract}

\section{Introduction}

Ninety years after Hubble's seminal work \citep{Hubble_1929}, observational cosmology remains focused on obtaining a precise and accurate value of the Hubble constant, $H_0$.  Today, measurements of the cosmic microwave background (CMB) at high redshift ($z \approx 1100$) determine the angular-size distance to the surface of last scattering and, within the context of the standard $\Lambda$CDM cosmological model, predict a precise value for $H_0$ of $67.4 \pm 0.5$\,\kmsmpc \citep{Planck_2018}.  Because this ``early-Universe'' prediction is model-dependent, complementary ``late-Universe'' measurements of $H_0$ provide an important test of the assumed cosmological model \citep[e.g.,][]{Hu_2005}.

Currently, measurements of $H_0$ at low redshifts ($z \ll 10$) are in statistical tension with the early-Universe prediction.  For example, distance ladder measurements using Cepheid variables to calibrate the absolute luminosities of Type Ia supernovae find $74.03 \pm 1.42$\,\kmsmpc \citep{Riess_2019}, and time-delay strong lensing measurements from multiply-imaged quasars currently yield $73.3^{+1.7}_{-1.8}$\,\kmsmpc \citep{Wong_2019}.  Though this discrepancy between the early- and late-Universe $H_0$ measurements is being taken increasingly seriously by the cosmological community \citep[see][]{Verde_2019}, the tantalizing prospects that it holds for heralding physics beyond $\Lambda$CDM \citep[e.g.,][]{Raveri_2017,Poulin_2018} require that we subject it to a correspondingly strict evidence threshold.  Independent avenues for constraining $H_0$ are thus necessary to provide cross-checks against unrecognized systematics in the measurements.

Water megamasers residing in the accretion disks around supermassive black holes (SMBHs) in active galactic nuclei (AGN) provide a unique way to bypass the distance ladder and make one-step, geometric distance measurements to their host galaxies.  The archetypal AGN accretion disk megamaser system is located in the nearby (7.6\,Mpc; \citealt{Humphreys_2013,Reid_2019}) Seyfert 2 galaxy NGC 4258 \citep{Claussen_1984,Nakai_1993,Herrnstein_1999}.  Very long baseline interferometric (VLBI) observations reveal that the masers trace the accretion disk on sub-parsec scales, where the SMBH dominates the gravitational potential.  The masing gas parcels act as test particles in this potential and exhibit the ordered, Keplerian motion expected for orbits about a point mass \citep{Greenhill_1995,Miyoshi_1995}.  By combining the VLBI position and velocity information with centripetal accelerations measured from spectral monitoring observations \citep[e.g.,][]{Argon_2007}, the typical degeneracy between mass and distance is broken and precise constraints can be placed on both quantities \citep[e.g.,][]{Herrnstein_1999,Humphreys_2013,Reid_2019}.

The Megamaser Cosmology Project (MCP) is a multi-year campaign to find, monitor, and map AGN accretion disk megamaser systems \citep{Braatz_2007,Braatz_2008b}.  The primary goal of the MCP is to constrain $H_0$ to a precision of several percent by making geometric distance measurements to megamaser galaxies in the Hubble flow \citep{Reid_2013,Kuo_2013,Kuo_2015,Gao_2017}.  Distance measurements made using the megamaser technique do not rely on distance ladders\footnote{Though maser-based distances are independent of standard candle distances, we note that the reverse is not always true.  There is one megamaser-hosting galaxy --  NGC 4258 -- whose distance measurement has been used to anchor standard candle luminosity calibrations, potentially resulting in correlated uncertainties.  However, \cite{Riess_2019} find that $H_0$ measurements made using only the Large Magellanic Cloud and Milky Way parallaxes as calibration anchors (i.e., excluding the NGC 4258 calibration) yield results that are consistent with those that include the NGC 4258 calibration.} or the CMB, and they have different systematics than lensing-based techniques.  The megamaser measurements thus provide an independent handle on $H_0$.

In this paper we present a revised analysis that improves the distance measurements for several megamaser systems that have been previously published by the MCP.  We then combine all distance measurements made using the revised analysis into a single megamaser-based constraint on $H_0$.  The paper is organized as follows.  In \autoref{sec:NewDistances} we discuss the new disk modeling and compare our revised distance estimates with the previously published results.  In \autoref{sec:Modeling} we describe how we combine the distance measurements into one $H_0$ constraint, detailing several different peculiar velocity treatments, and we present our resulting $H_0$ measurement.  We summarize and conclude in \autoref{sec:Conclusions}.

\section{Improved distance estimates} \label{sec:NewDistances}

In this section we present updated distance measurements for the megamaser-hosting galaxies UGC 3789, NGC 6264, NGC 6323, and NGC 5765b.  Each of these galaxies has had a maser-derived distance measurement published previously by the MCP \citep[see][]{Reid_2013,Kuo_2013,Kuo_2015,Gao_2017}.

The improvements we present here stem primarily from an update to the fitting procedure that incorporates the ``error floor'' systematic uncertainties as model parameters, thereby enabling marginalization over a previous source of systematic uncertainty.  These error floors get added in quadrature with the data errors, such that an error floor value of zero indicates that the data errors already capture the true measurement uncertainties well.  This updated model has already been applied to the galaxies NGC 4258 \citep{Reid_2019} and CGCG 074-064 \citep{Pesce_2020}, so for these galaxies we simply use the corresponding published results.

We use a Hamiltonian Monte Carlo (HMC) sampler as implemented within the PyMC3 package \citep{Salvatier2016} to perform all of the disk fitting described in this section; for a comprehensive overview of the disk model and fitting procedure, see \cite{Pesce_2020}.  The HMC code yields results that match well with those from the Metropolis-Hastings disk-fitting code employed in many previous MCP publications \citep[see, e.g.,][]{Reid_2013}, but has an improved convergence efficiency.  We use broad truncated Gaussian priors for all error floor parameters, with a mean of 10\,$\mu$as and a standard deviation of 5\,$\mu$as for both the $x$- and $y$-position error floor priors, a mean of 2\,\kms and a standard deviation of 1\,\kms for both the systemic and high-velocity\footnote{Disk maser spectra are typically characterized by three groups of maser features arranged roughly symmetrically about the systemic velocity of the galaxy \citep[see, e.g.,][]{Nakai_1993,Braatz_2004}, corresponding to the locations in an edge-on accretion disk where the line-of-sight velocity gradient is minimized.  The ``systemic masers'' sit in front of the black hole as seen along the line-of-sight, and in spectra they appear approximately centered on the systemic velocity of the system.  The ``high-velocity masers'' originate on the ``midline'' of the disk (i.e., where the sky plane intersects the plane of the disk) where the line-of-sight passes tangent to the orbital velocity vectors, and in spectra they appear Doppler shifted by typically several hundred \kms to either side of the systemic velocity.} error floor priors, and a mean of 0.3\,\kmsyr and a standard deviation of 0.15\,\kmsyr for the acceleration error floor priors.  In all cases, the prior distribution is truncated at zero so that the error floors remain strictly positive.

\subsection{UGC 3789} \label{sec:UGC3789}

The disk megamaser in the galaxy UGC 3789 was discovered by \cite{Braatz_2008a}, and the first VLBI map of the system was presented in \cite{Reid_2009}.  This map was combined with acceleration measurements by \cite{Braatz_2010} to produce an angular-size distance estimate of $49.9 \pm 7.0$\,Mpc to the system.  \cite{Reid_2013} obtained additional data and improved the distance measurement to $49.6 \pm 5.1$\,Mpc, and we use the data from that latest paper to produce the updated measurements presented in \autoref{tab:UpdatedDiskFitting} and \autoref{app:DiskModelingResults}.

Our updated fit indicates that the error floors used by \cite{Reid_2013} for the position measurements were too conservative; we find a best-fit $x$-position error floor of $5 \pm 1$\,$\mu$as and a best-fit $y$-position error floor of $6 \pm 1$\,$\mu$as, compared to the previous (fixed) values of 10\,$\mu$as for both $x$- and $y$-positions.  We find that the data do not provide strong constraints on the velocity measurement error floors for either the systemic or the high-velocity masers, and that the posteriors for both largely follow the Gaussian-distributed prior range of $2.0 \pm 1.0$\,\kms.  Our best-fit acceleration error floor is $0.34 \pm 0.06$\,\kmsyr, again indicating that the \cite{Reid_2013} value of 0.57\,\kmsyr was too conservative.

The net result of the updated modeling is a distance measurement of $D = 51.5^{+4.5}_{-4.0}$\,Mpc, which represents a $3.8\% \approx 0.4{\sigma}$ increase over the previous value and an improvement in the measurement precision from $\pm 10\%$ to $(+8.7\%,-7.8\%)$.

\begin{deluxetable*}{lCCc}[t]
\tablecolumns{4}
\tablewidth{0pt}
\tablecaption{Results from maser disk modeling \label{tab:UpdatedDiskFitting}}
\tablehead{\colhead{Galaxy} & \colhead{Distance (Mpc)} & \colhead{Velocity (\kms)} & \colhead{Reference}}
\startdata
UGC 3789     & 51.5^{+4.5}_{-4.0}   & \hphantom{0}3319.9 \pm 0.8  & this work \\
NGC 6264     & 132.1^{+21}_{-17}    & 10192.6 \pm 0.8  & this work \\
NGC 6323     & 109.4^{+34}_{-23}    & \hphantom{0}7801.5 \pm 1.5  & this work \\
NGC 5765b    & 112.2^{+5.4}_{-5.1}  & \hphantom{0}8525.7 \pm 0.7  & this work \\
CGCG 074-064 & 87.6^{+7.9}_{-7.2}   & \hphantom{0}7172.2 \pm 1.9  & \cite{Pesce_2020} \\
NGC 4258     & 7.58 \pm 0.11        & \hphantom{00}679.3 \pm 0.4  & \cite{Reid_2019} \\
\enddata
\tablecomments{Maser galaxy distances and velocities as measured from modeling the maser disks; for each value we quote the posterior median and 1$\sigma$ confidence interval (i.e., 16th to 84th percentile).  For NGC 4258, where the systematic uncertainty in the distance measurement is comparable to its statistical uncertainty, we have added the two in quadrature.  All velocities are quoted in the CMB reference frame using the optical convention.  The values for the other parameters measured from the disk model are given in \autoref{app:DiskModelingResults}.}
\end{deluxetable*}

\subsection{NGC 6264} \label{sec:NGC6264}

A VLBI map for the maser system in NGC 6264 was first presented in \cite{Kuo_2011}, and \cite{Kuo_2013} reported an angular-size distance measurement of $144 \pm 19$\,Mpc.  We use the data from the latter paper to produce the updated measurements in \autoref{tab:UpdatedDiskFitting} and \autoref{app:DiskModelingResults}.

\cite{Kuo_2013} imposed error floors of 8\,$\mu$as on the $x$-position measurements and 16\,$\mu$as on the $y$-position measurements.  We find that both of these values were too conservative; the updated fitting prefers $x$-position error floors that are consistent with zero and $y$-position error floors of $6 \pm 2$\,$\mu$as.  Our best-fit velocity error floors are only moderately constrained by the data (beyond the prior constraints); we find values of $1.6^{+0.9}_{-0.8}$\,\kms and $1.3^{+0.7}_{-0.6}$\,\kms for the systemic and high-velocity features, respectively, consistent with the error floors used in the previous disk modeling.  We find that the acceleration error floor is consistent with zero and confined to a range of values that is considerably smaller than the $\sim$0.3--0.7\,\kmsyr imposed on the measurements by \cite{Kuo_2013}.

The updated modeling yields an angular-size distance measurement of $D = 132.1^{+21}_{-17}$\,Mpc, representing a $9\% \approx 0.6{\sigma}$ decrease compared to the previously-published value.  The uncertainty in the distance measurement remains essentially unchanged.

\subsection{NGC 6323} \label{sec:NGC6323}

The maser system in the galaxy NGC 6323 was discovered by \cite{Braatz_2004}, and the first VLBI map was presented in \cite{Braatz_2007}.  \cite{Kuo_2015} combined additional epochs of VLBI with spectral monitoring observations and reported an angular-size distance measurement of $107^{+42}_{-27}$\,Mpc.  We use the data from the latter paper to produce the updated measurements in \autoref{tab:UpdatedDiskFitting} and \autoref{app:DiskModelingResults}.

The original error floors in the position measurements were set to 10\,$\mu$as, while we find that the updated disk modeling prefers a smaller value of $4 \pm 1$\,$\mu$as for the $x$-positions and is consistent with zero for the $y$-positions.  We find that the data aren't able to place constraints on the velocity error floors, as our posteriors recover the prior Gaussian distributions of $2 \pm 1$\,\kms.  Our best-fit acceleration error floor is consistent with zero.

Our updated disk modeling constrains the distance to be $D = 109^{+34}_{-23}$\,Mpc.  This value matches well with the $107^{+42}_{-29}$\,Mpc distance reported in \cite{Kuo_2015}, and we have improved the measurement precision from $(+39\%,-27\%)$ to $(+31\%,-21\%)$.

\subsection{NGC 5765b} \label{sec:NGC5765b}

\cite{Gao_2017} presented the maser system in NGC 5765b and measured its angular-size distance to be $122.0^{+10.0}_{-8.6}$\,Mpc.\footnote{\cite{Gao_2017} reported a Hubble constant of $66.0 \pm 5.0$\,Mpc using a recession velocity of 8334.6\,\kms, which corresponds to a distance of $126.3^{+10.3}_{-8.9}$\,Mpc using $D = v / H_0$.  For this paper we convert between $D$ and $H_0$ using \autoref{eqn:ModelDistance}, so the distance of $122.0^{+10.0}_{-8.6}$\,Mpc we mention here differs slightly from that reported in \cite{Gao_2017}.}  We use the data from that paper to produce the updated measurements in \autoref{tab:UpdatedDiskFitting} and \autoref{app:DiskModelingResults}.

We find that our disk fit prefers error floors of $3 \pm 1$\,$\mu$as for both the $x$- and $y$-position data, indicating that the $10$\,$\mu$as error floors used in \cite{Gao_2017} were too conservative.  Conversely, we find the 0.6\,\kms velocity error floors used in the original disk modeling were too optimistic, and the data place modest constraints on the velocity error floors of $1.1 \pm 0.5$\,\kms and $1.5 \pm 0.5$\,\kms for the systemic and high-velocity features, respectively.  Our best-fit acceleration error floor of $0.04 \pm 0.01$\,\kmsyr is on the low end of the 0.05--0.2\,\kmsyr range used in \cite{Gao_2017}.

Our updated disk modeling yields a distance measurement of $D = 112.2^{+5.4}_{-5.1}$\,Mpc, a $9\% \approx 1{\sigma}$ decrease compared to the previously-published value.  The distance uncertainty has improved from $(+8.2\%,-7.0\%)$ to $(+4.8\%,-4.5\%)$.

\begin{figure*}
    \centering
    \includegraphics[width=0.56\textwidth]{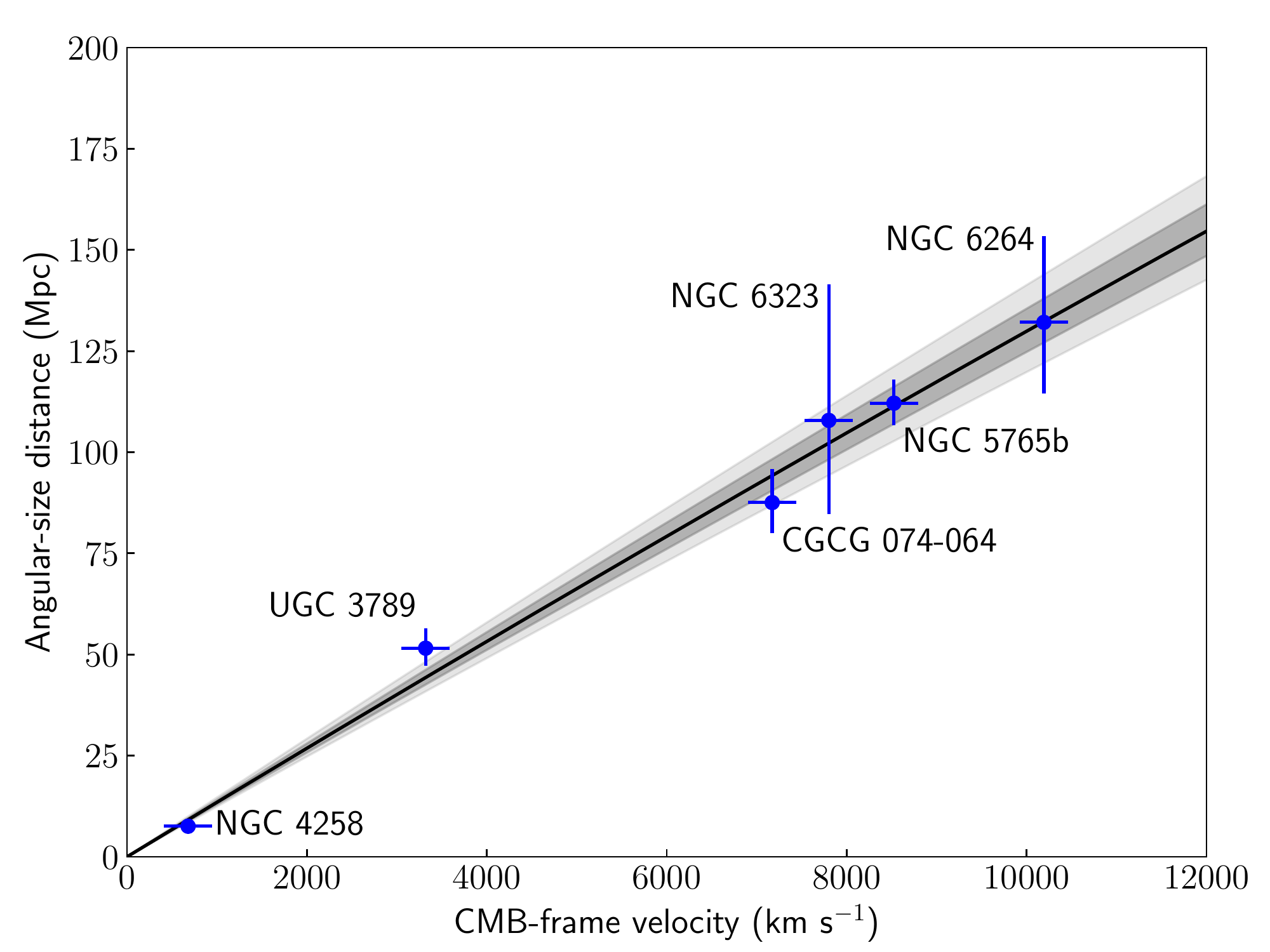}
    \includegraphics[width=0.42\textwidth]{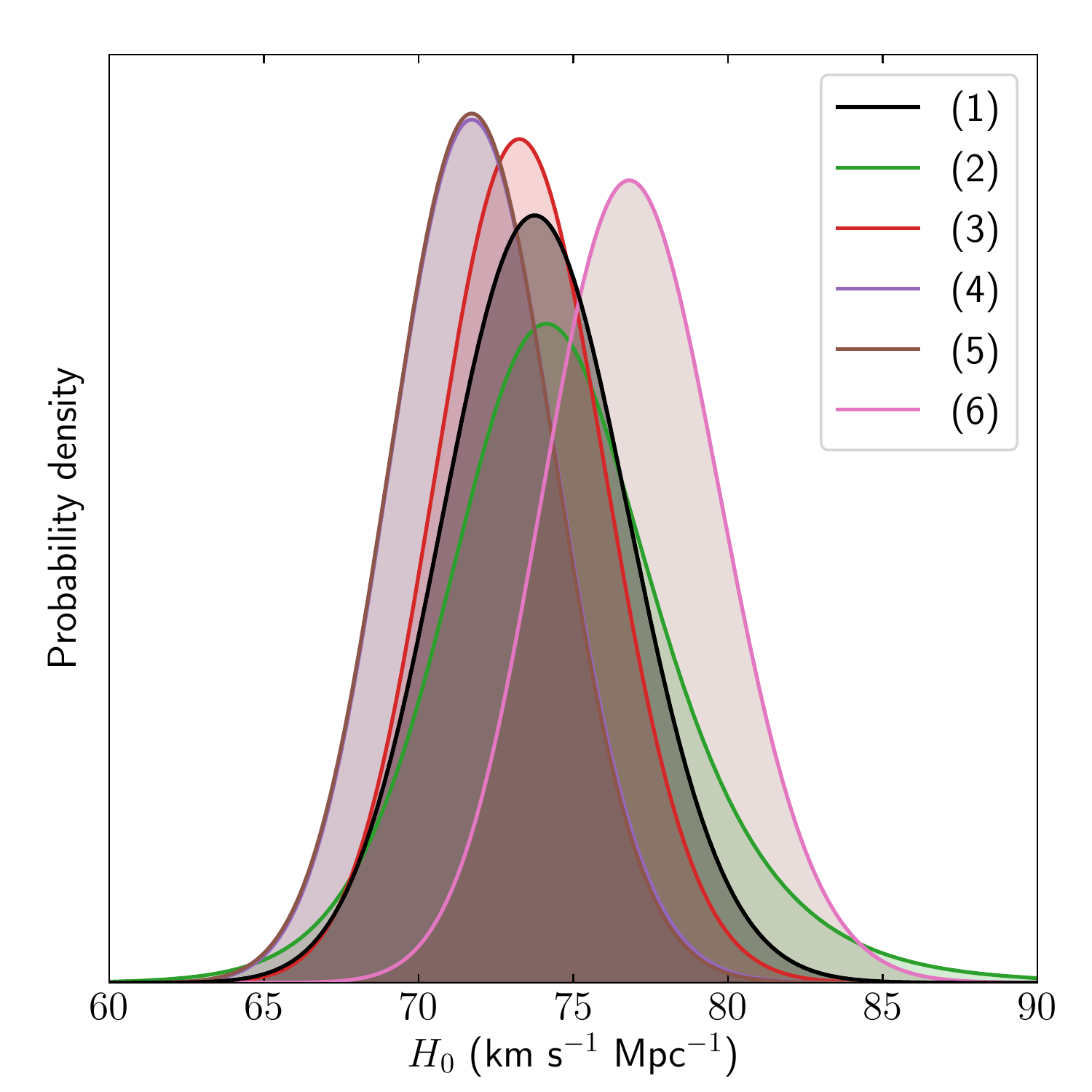}
    \caption{\textit{Left}: Hubble diagram for the maser galaxies considered in this paper.  Each data point is plotted with $1{\sigma}$ uncertainties in distance and 250\,\kms uncertainties in velocity.  The solid black line shows the distance-velocity relationship from \autoref{eqn:ModelDistance} for the maximum-likelihood $H_0$ value corresponding to the peculiar velocity treatment described in \autoref{sec:PeculiarUncertainties}, and the shaded gray regions show $1{\sigma}$ and $2{\sigma}$ confidence intervals.  \textit{Right}: Posterior distributions for $H_0$ from the five different peculiar velocity treatments considered in this paper; the treatments are numbered as in \autoref{tab:HubbleConstants}.  Our ``fiducial'' treatment is (1) and is plotted in black.  Note that treatments (4) and (5), corresponding to the galaxy flow corrections using 2M++ and CF3, return nearly identical $H_0$ posteriors.}
    \label{fig:HubbleDiagram}
\end{figure*}

\begin{deluxetable*}{lcC}[t]
\tablecolumns{3}
\tablewidth{0pt}
\tablecaption{Hubble constant constraints and jackknife tests \label{tab:HubbleConstants}}
\tablehead{\colhead{Peculiar velocity treatment} & \colhead{Galaxies excluded from the fit} & \colhead{$H_0$ (\kmsmpc)}}
\startdata
\multirow{7}{8cm}{(1) Assign a fixed velocity uncertainty of 250\,\kms}    & UGC 3789      & 75.8^{+3.4}_{-3.3}  \\
    & NGC 6264      & 73.8^{+3.2}_{-3.2}  \\
    & NGC 6323      & 73.8^{+3.1}_{-3.0} \\
    & NGC 5765b     & 74.1^{+4.5}_{-4.4}  \\
    & CGCG 074-064   & 72.5^{+3.4}_{-3.2}  \\
    & NGC 4258      & 73.6^{+3.1}_{-3.0}  \\
\cmidrule{2-3}
    & \textbf{Fit using all galaxies:}          & 73.9^{+3.0}_{-3.0} \\
\midrule
\multirow{7}{8cm}{(2) Fit for $\sigma_{\text{pec}}$ using the maser data and assuming an outlier-robust form for the peculiar velocity distribution}    & UGC 3789      & 76.4^{+4.2}_{-3.8} \\
    & NGC 6264      &  74.4^{+4.4}_{-3.8} \\
    & NGC 6323      & 74.5^{+4.0}_{-3.6} \\
    & NGC 5765b     & 75.8^{+6.6}_{-5.6}  \\
    & CGCG 074-064   & 73.1^{+4.3}_{-3.9} \\
    & NGC 4258      & 74.2^{+4.5}_{-3.7} \\
\cmidrule{2-3}
    & \textbf{Fit using all galaxies:}          & 74.4^{+3.9}_{-3.4} \\
\midrule
\multirow{7}{8cm}{(3) Use galaxy group recession velocities}    & UGC 3789      & 75.0^{+3.1}_{-3.0}  \\
    & NGC 6264      & 73.1^{+2.8}_{-2.7}  \\
    & NGC 6323      & 73.2^{+2.8}_{-2.7}  \\
    & NGC 5765b     & 72.2^{+4.2}_{-4.1}  \\
    & CGCG 074-064   & 73.3^{+3.1}_{-3.0}  \\
    & NGC 4258      & 72.8^{+2.8}_{-2.7}  \\
\cmidrule{2-3}
    & \textbf{Fit using all galaxies:}          & 73.3^{+2.8}_{-2.7} \\
\midrule
\multirow{7}{*}{(4) Use 2M++ \citep{Carrick_2015} recession velocities}    & UGC 3789      & 73.3^{+3.0}_{-3.0}  \\
    & NGC 6264      & 71.8^{+2.8}_{-2.8}  \\
    & NGC 6323      & 71.9^{+2.8}_{-2.7}  \\
    & NGC 5765b  & 71.1^{+4.0}_{-3.9}  \\
    & CGCG 074-064   & 70.9^{+3.0}_{-2.9}  \\
    & NGC 4258      & 72.1^{+2.7}_{-2.7}  \\
\cmidrule{2-3}
    & \textbf{Fit using all galaxies:}          & 71.8^{+2.7}_{-2.7} \\
\midrule
\multirow{7}{8cm}{(5) Use CF3 \citep{Graziani_2019} recession velocities}    & UGC 3789      & 73.6^{+3.1}_{-2.9}   \\
    & NGC 6264      & 71.5^{+2.8}_{-2.7}   \\
    & NGC 6323      & 71.7^{+2.8}_{-2.6}   \\
    & NGC 5765b     & 71.5^{+4.1}_{-4.0}   \\
    & CGCG 074-064   & 70.5^{+3.0}_{-2.9}   \\
    & NGC 4258      &  72.0^{+2.7}_{-2.7}  \\
\cmidrule{2-3}
    & \textbf{Fit using all galaxies:}          & 71.8^{+2.7}_{-2.6} \\
\midrule
\multirow{7}{8cm}{(6) Use M2000 \citep{Mould_2000} recession velocities}    & UGC 3789      & 79.3^{+3.3}_{-3.1}   \\
    & NGC 6264      & 76.8^{+2.9}_{-2.9}   \\
    & NGC 6323      & 76.9^{+2.9}_{-2.9}   \\
    & NGC 5765b     & 76.2^{+4.3}_{-4.1}   \\
    & CGCG 074-064   & 75.5^{+3.2}_{-3.0}   \\
    & NGC 4258      &  76.8^{+2.9}_{-2.9}  \\
\cmidrule{2-3}
    & \textbf{Fit using all galaxies:}          & 76.9^{+2.9}_{-2.9} \\
\enddata
\tablecomments{Hubble constant measurements made using various subsets of the megamaser distances and different treatments for the peculiar velocities, as described in \autoref{sec:Modeling}.  For each peculiar velocity treatment, we list seven $H_0$ values: six of these values correspond to ``leave-one-out'' jackknife tests, in which we fit the data (under the given peculiar velocity prescription) after removing the galaxy specified in the second column; the seventh value corresponds to that obtained from fitting all galaxies simultaneously. For each fit we quote the posterior median and 1$\sigma$ confidence interval (i.e., 16th to 84th percentile).}
\end{deluxetable*}

\begin{deluxetable*}{lCCC}[t]
\tablecolumns{4}
\tablewidth{0pt}
\tablecaption{$H_0$ goodness-of-fits and comparisons with other measurements \label{tab:FitStatistics}}
\tablehead{\colhead{Peculiar velocity treatment} & \colhead{$\chi_{\nu}^2$} & \colhead{$P(H_0 \leq H_{0,\text{Planck}})$} & \colhead{$P(H_0 \geq H_{0,\text{SH0ES}})$}}
\startdata
(1) & 0.60 & 0.02 & 0.48 \\
(2) & 1.52 & 0.03 & 0.55 \\
(3) & 0.62 & 0.01 & 0.41 \\
(4) & 0.55 & 0.05 & 0.24 \\
(5) & 0.75 & 0.05 & 0.23 \\
(6) & 0.75 & <0.01\hphantom{<} & 0.82 \\
\enddata
\tablecomments{Statistics for the Hubble constant fits described in \autoref{sec:Modeling}, with the different peculiar velocity treatments numbered as in \autoref{tab:HubbleConstants}.  The second column lists the chi-squared per degree of freedom for each fit, computed using \autoref{eqn:ChiSquared}.  For treatments (1), (3), (4), (5), and (6), the number of degrees of freedom $\nu = 5$ and the expected standard deviation of the $\chi_{\nu}^2$-distribution is $\sqrt{2/5} \approx 0.63$, while for treatment (2), $\nu = 4$ and the expected standard deviation in $\chi_{\nu}^2$ is $\sqrt{2/4} \approx 0.71$.  The third column lists one-sided comparison statistics computed using \autoref{eqn:PlanckComparison}, which give the probability that our $H_0$ measurement is at least as low as the Planck measurement \citep{Planck_2018}.  The fourth column is analogous to the third, and lists the probability that our $H_0$ measurement is at least as high as the SH0ES measurement (\citealt{Riess_2019}; the statistic is computed using \autoref{eqn:SH0ESComparison}).}
\end{deluxetable*}

\section{Modeling the Hubble constant} \label{sec:Modeling}

Our disk model returns an angular-size distance measurement $\hat{D}_i$ and a redshift measurement $\hat{z}_i$ for the SMBH in each megamaser-hosting galaxy.  From considerations of kinetic energy equipartition between the SMBH and surrounding stars, we expect the relative velocity of a ${\sim}10^7$\,M$_{\odot}$ black hole to be $\ll$1\,\kms with respect to the system barycenter \citep{Merritt_2007}.  Measured upper limits on the magnitude of this relative motion are on the order of several \kms for the SMBH in the center of the Milky Way \citep{Reid_2004,Reid_2020} and no more than a few tens of \kms for the sources considered in this work \citep{Pesce_2018}.  We thus proceed under the assumption that each galaxy effectively shares a distance and redshift with its SMBH, and we seek to determine what values of $H_0$ are compatible with these measurements.

For each galaxy, its expected angular-size distance $D_i$ is related to its expected cosmological recession redshift $z_i$ and $H_0$ by

\begin{eqnarray}
D_i & = & \frac{c}{H_0 \left( 1 + z_i \right)} \int_0^{z_i} \frac{\text{d}z}{\sqrt{\Omega_m \left( 1 + z \right)^3 + \left( 1 - \Omega_m \right)}} \nonumber \\
& \approx & \frac{c z_i}{H_0 \left( 1 + z_i \right)} \left( 1 - \frac{3 \Omega_m z_i}{4} + \frac{\Omega_m ( 9 \Omega_m - 4) z_i^2}{8} \right) , \label{eqn:ModelDistance}
\end{eqnarray}

\noindent where by ``expected'' here we refer to the values that the distance and redshift would take if the galaxy were perfectly following the Hubble flow.  \autoref{eqn:ModelDistance} assumes a flat $\Lambda$CDM cosmology, and the series expansion is accurate to one part in ${\sim}10^5$ for the range of redshifts covered by our observations.  We set $\Omega_m = 0.315$, from \cite{Planck_2018}, though we note that any choice in the range $0 \leq \Omega_m \leq 0.5$ would yield a distance that differs by ${\lesssim}1\%$ for the galaxies in our sample.  For each of the $H_0$-fitting approaches described in this section, the contribution to the likelihood from the distance constraints is given by the product of the posterior distributions $\mathcal{P}(\hat{D}_i | D_i)$ from the independent disk fits,

\begin{equation}
\mathcal{L}_D = \prod_i \mathcal{P}(\hat{D}_i | D_i) ,
\end{equation}

\noindent where $\hat{D}_i$ is the distance measured from disk modeling.  

The expected cosmological recession velocity $v_i = c z_i$ differs from the measured galaxy velocity $\hat{v}_i$ both because of statistical uncertainty in the measurement and because of a systematic uncertainty in the form of peculiar motion.  For our measurements, in which the statistical uncertainties in velocity are quite small (typically $\sim$1--2\,\kms), peculiar motions dominate the recession velocity uncertainty.  Because the galaxies in our sample all reside at low redshifts ($z \ll 1$), we proceed under the assumption that peculiar velocities are independent of redshift.

When fitting for $H_0$ we have several options for treating these peculiar velocities, and in this section we describe the approaches we have taken to construct the velocity contribution to the likelihood, $\mathcal{L}_v$.  In all cases, our combined likelihood $\mathcal{L}$ is ultimately given by the product of the velocity and distance likelihoods,

\begin{equation}
\mathcal{L} = \mathcal{L}_D \mathcal{L}_v ,
\end{equation}

\noindent and the posterior distribution is given by the product of $\mathcal{L}$ with the prior via Bayes's theorem.  We assume flat priors for all model parameters, and we explore the posterior space using the \texttt{dynesty} nested sampling package \citep{Speagle_2019}.

\subsection{Treating peculiar velocities as inflated measurement uncertainties} \label{sec:PeculiarUncertainties}

The simplest way to take peculiar velocities into account is to incorporate them into the velocity measurement uncertainties.  Typical values for galaxy peculiar velocities lie in the range ${\sim}150$--250\,\kms \citep[e.g.,][]{Davis_1997,Zaroubi_2001,Masters_2006,Hoffman_2015}, so we conservatively take the upper end of this range and add $\sigma_{\text{pec}} = 250$\,\kms in quadrature to our velocity uncertainties.  The velocity contribution to the likelihood is then given by a Gaussian distribution,

\begin{equation}
\mathcal{L}_v = \prod_i \frac{1}{\sqrt{2 \pi \left( \sigma_{v,i}^2 + \sigma_{\text{pec}}^2 \right)}} \exp\left( - \frac{1}{2} \frac{\left( v_i - \hat{v}_i \right)^2}{\sigma_{v,i}^2 + \sigma_{\text{pec}}^2} \right) , \label{eqn:VelocityLikelihood1}
\end{equation}

\noindent where $\sigma_{v,i}$ is the statistical uncertainty in velocity measurement $\hat{v}_i$ and the true velocities $v_i$ are treated as nuisance parameters in the model.

The result from fitting this model to all six maser galaxies simultaneously is $H_0 =  73.9 \pm 3.0$\,\kmsmpc, and \autoref{tab:HubbleConstants} lists the values obtained from leave-one-out jackknife tests.  We assess the goodness-of-fit using a chi-squared statistic,

\begin{equation}
\chi_{\nu}^2 = \frac{1}{\nu} \sum_i \left[ \frac{\left( v_i - \hat{v}_i \right)^2}{\sigma_{v,i}^2 + \sigma_{\text{pec}}^2} + \frac{( D_i - \hat{D}_i )^2}{\sigma_{D,i}^2} \right] , \label{eqn:ChiSquared}
\end{equation}

\noindent where $\sigma_{D,i}$ is the standard deviation of the distance measurement posterior and $\chi_{\nu}^2$ is the chi-squared per degree of freedom, $\nu$.  For this fit, $\nu = 5$ and $\chi_{\nu}^2 = 0.6$, which is consistent with unity within the expected standard deviation of a chi-squared distribution with five degrees of freedom (see also \autoref{tab:FitStatistics}).

\subsection{Modeling peculiar velocities as being drawn from a global distribution} \label{sec:GlobalPeculiar}

Rather than assuming a typical dispersion for the peculiar velocity distribution of $\sigma_{\text{pec}}$, we can instead fit for it as part of the model given some assumption about the form of the underlying distribution from which peculiar velocities are drawn.  Because we expect that some unknown fraction of galaxies may have particularly large peculiar velocities (e.g., if the galaxy lives in a cluster), we test the \citet[][\S 8.3.1]{Sivia_2006} ``conservative formulation'' for the velocity uncertainties as an alternative to a Gaussian distribution.  Under this formalism, $\sigma_{\text{pec}}$ is interpreted as a lower bound on the velocity error $\sigma$ associated with peculiar velocities, with a distribution for this error given by

\begin{equation}
\mathcal{P}(\sigma | \sigma_{\text{pec}}) = \begin{cases}
\frac{\sigma_{\text{pec}}}{\sigma^2} , & \sigma > \sigma_{\text{pec}} \\
0 , & \text{otherwise}
\end{cases} .
\end{equation}

\noindent The marginal likelihood contribution from the velocity constraints after integrating out $\sigma$ is then

\begin{equation}
\mathcal{L}_v = \prod_i \frac{1}{\sqrt{2 \pi \left( \sigma_{v,i}^2 + \sigma_{\text{pec}}^2 \right)}} \left( \frac{1 - e^{-R_i^2/2}}{R_i^2} \right) , \label{eqn:ConservativeLikelihood}
\end{equation}

\noindent where

\begin{equation}
R_i = \frac{v_{\text{pec},i}^2}{\sigma_{v,i}^2 + \sigma_{\text{pec}}^2} 
\end{equation}

\noindent as we have assumed that the peculiar velocity distribution has zero mean.

Because we are now modeling peculiar velocity rather than recession velocity, we have to relate the two.  We relate the expected cosmological recession redshift $z_i$ to the measured redshift $\hat{z}_i$ by adding the peculiar velocity contribution via \citep[see, e.g.,][]{Davis_2014},

\begin{equation}
1 + \hat{z}_i = \left( 1 + z_i \right) \left( 1 + \frac{v_{\text{pec},i}}{c} \right) . \label{eqn:ModelRedshift}
\end{equation}

\noindent The $z_i$ values are then plugged into \autoref{eqn:ModelDistance} to compute the distances $D_i$.

The result from fitting this model to all six maser galaxies simultaneously is $H_0 = 74.4^{+3.9}_{-3.4}$\,\kmsmpc, and \autoref{tab:HubbleConstants} lists the values obtained from leave-one-out jackknife tests.  We find $\sigma_{\text{pec}} = 141^{+185}_{-80}$\,\kms and use it via \autoref{eqn:ChiSquared} to compute $\chi_{\nu}^2 = 1.5$ (see \autoref{tab:FitStatistics}), which is consistent with unity within the expected standard deviation for a chi-squared distribution with $\nu = 4$ degrees of freedom.  We note also that the explicitly non-Gaussian form of the likelihood in \autoref{eqn:ConservativeLikelihood} -- in particular the $1/R_i^2$ tails of this distribution -- are expected to drive the chi-squared above unity.

\subsection{Using galaxy group velocities in place of individual galaxy velocities} \label{sec:GroupCorrection}

As an alternative to treating the peculiar velocity as an uncertainty in the measured velocity, we can use external information to independently estimate the recession velocity for each galaxy in our sample.  One of the primary drivers of peculiar motion is dispersion within galaxy groups or clusters, so we can attempt to correct for such dispersion by associating the cosmological recession velocity for each galaxy with the velocity of the group that galaxy resides in.

We use the galaxy groups defined in \cite{Tully_2015}, and we take the velocities for the galaxies within each group from NED.\footnote{The NASA/IPAC Extragalactic Database (NED) is funded by the National Aeronautics and Space Administration and operated by the California Institute of Technology.}  We define the group velocity to be the mean of all galaxy velocities within that group; these group recession velocities are listed in \autoref{tab:VariousVels} along with their corresponding peculiar velocity equivalents.  We note that not all of these group associations are equally certain, and that in particular the large peculiar velocity predicted for CGCG 074-064 (which was not associated with a galaxy group by \citealt{Lavaux_2011}) may indicate that it warrants further investigation.

Given a recession velocity $\hat{v}_i$ for each galaxy group, we fit the model in the same manner described in \autoref{sec:PeculiarUncertainties} except that we apply a 150\,\kms rather than a 250\,\kms global velocity uncertainty.  We choose the low end of the plausible peculiar velocity range (see \autoref{sec:PeculiarUncertainties}) because the galaxy groups, being much more massive than individual galaxies, should exhibit a smaller dispersion about the Hubble flow.

Fitting this model to all six maser galaxies simultaneously, we find $H_0 = 73.3^{+2.8}_{-2.7}$\,\kmsmpc.  \autoref{tab:HubbleConstants} lists the values obtained from leave-one-out jackknife tests for both models.  We compute $\chi_{\nu}^2$ using \autoref{eqn:ChiSquared} (see \autoref{tab:FitStatistics}), and we find that the chi-squared values are consistent with unity within the expected standard deviation of a chi-squared distribution with $\nu = 5$ degrees of freedom.

\subsection{Using galaxy flow models to correct for peculiar velocities} \label{sec:FlowCorrection}

The last peculiar velocity treatment we consider is similar to the one described in the previous section in that it relies on the use of external information to constrain the permitted velocities for each galaxy in our sample.  We obtain peculiar velocity predictions from three different galaxy flow models: (1) the ``2M++ model'' constructed by \cite{Carrick_2015} using the 2M++ redshift catalog \citep{Lavaux_2011}; (2) the ``CF3 model'' constructed by \cite{Graziani_2019} from the \textit{Cosmicflows-3} extragalactic distance database \citep{Tully_2016}; and (3) the ``M2000 model'' constructed by \cite{Mould_2000} to model the impact of the Great Attractor (GA), Virgo Cluster, and Shapley Cluster on galaxy motions.  Each of these catalogs makes a prediction for the recession velocity associated with a particular sky direction and redshift, and we list these velocities in \autoref{tab:VariousVels}.

Given the recession velocity predictions from the catalogs, the model is fit in an identical manner to that used in \autoref{sec:PeculiarUncertainties} and \autoref{sec:GroupCorrection}.  We assume a 150\,\kms uncertainty for all recession velocities, as per \citet{Carrick_2015} for 2M++ and \citet{Graziani_2019} for CF3.

Though each galaxy flow model makes different predictions for individual galaxy recession velocities (see \autoref{tab:VariousVels}), the results from fitting the 2M++ and CF3 models to all six maser galaxies simultaneously are in good agreement.  We find $H_0 = 71.8 \pm 2.9$\,\kmsmpc when using the 2M++ velocities and $H_0 = 71.8^{+2.7}_{-2.6}$\,\kmsmpc for the CF3 velocities, both somewhat lower values than found using the treatments in previous sections.

For the M2000 model, we instead find a larger best-fit Hubble constant of $H_0 = 76.9 \pm 2.9$\,\kmsmpc.  The M2000 model is simpler than either 2M++ or CF3 in that it considers only the gravitational influences of three large structures on individual galaxy motions, rather than using a global density field as done in both 2M++ and CF3.  The substantial difference between the $H_0$ value predicted by the M2000 model and that predicted by the 2M++/CF3 models results from the effect of the GA in the M2000 infall model on three of the maser host galaxies.  In the M2000 model, NGC 6264, NGC 5765b, and CGCG 074-064 are all in the direction of the GA and beyond it, reducing their perceived redshifts by our and their infall into the GA by 300--600\,\kms.  These values get added back to the host galaxy velocities and thereby increase $H_0$.  The veracity of these corrections is thus sensitive to the position and scale of the GA, which remains a subject of some debate \citep{Dressler_1987,Tully_2014,Kraan-Korteweg_2016,Hoffman_2017}

\autoref{tab:HubbleConstants} lists the $H_0$ values obtained from leave-one-out jackknife tests for each of the three models considered in this section, and \autoref{tab:FitStatistics} lists the $\chi_{\nu}^2$ values computed using \autoref{eqn:ChiSquared}.  In all three cases we find that the $\chi_{\nu}^2$ values are consistent with unity within the expected standard deviation of a chi-squared distribution with $\nu=5$ degrees of freedom.

\begin{deluxetable*}{lCCCCCCCCCC}[t]
\tablecolumns{11}
\tablewidth{0pt}
\tablecaption{Relevant velocities for maser galaxies \label{tab:VariousVels}}
\tablehead{ & & \multicolumn{4}{c}{Predicted recession velocity (\kms)} & & \multicolumn{4}{c}{Peculiar velocity (\kms)} \\ \cline{3-6} \cline{8-11}
\colhead{Galaxy} & \colhead{Velocity (\kms)} & \colhead{Group} & \colhead{2M++} & \colhead{CF3} & \colhead{M2000} &  & \colhead{Group} & \colhead{2M++} & \colhead{CF3} & \colhead{M2000}}
\startdata
UGC 3789     & \hphantom{0}3319.9   & \hphantom{0}3401 & \hphantom{0}3375 & 3292  & 3464 &  & -80   & -54 & \hphantom{0}28 & -142 \\
NGC 6264     & 10192.6              & 10379 & \hphantom{0}9962 & 10175\hphantom{0} & 10677 &  & -180\hphantom{0}   & \hphantom{-}224\hphantom{0} & \hphantom{0}17 & -468 \\
NGC 6323     & \hphantom{0}7801.5   & \hphantom{0}8275 & \hphantom{0}7378 & 8112 & 8208 &  & -461\hphantom{0} & \hphantom{-}414\hphantom{0} & -302\hphantom{-} & -396 \\
NGC 5765b    & \hphantom{0}8525.7   & \hphantom{0}8594 & \hphantom{0}8398 & 8333 & 9000 &  & -66              & \hphantom{-}124\hphantom{0} & 187 & -460 \\
CGCG 074-064 & \hphantom{0}7172.2   & \hphantom{0}6511 & \hphantom{0}6869 & 7037 & 7554 &  & \hphantom{-}647\hphantom{0}  & \hphantom{-}297\hphantom{0} & 132 & -372 \\
NGC 4258     & \hphantom{00}679.3   & \hphantom{00}725 & \hphantom{00}417 & \hphantom{0}425 & 581 &  & -46              & \hphantom{-}262\hphantom{0} & 254 & \hphantom{-0}98
\enddata
\tablecomments{Various velocities relevant for the maser galaxies considered in this paper.  The ``velocity'' column lists the CMB-frame velocity for the galaxy (defined such that $v = cz$) as measured from the maser disk fitting; we take this velocity to be a measure of the \textit{actual} redshift of the galaxy with respect to us.  The ``predicted recession velocity'' columns list the \textit{expected} CMB-frame recession velocities for each galaxy (again defined such that $v = cz$) from one of two specified treatments; the group treatment is described in \autoref{sec:GroupCorrection}, and the 2M++ \citep{Carrick_2015} / CF3 \citep{Graziani_2019} / M2000 \citep{Mould_2000} treatment is described in \autoref{sec:FlowCorrection}.  The CF3 recession velocities come courtesy of R. Graziani, private communication.  The ``peculiar velocity'' columns list the implied peculiar velocity for the galaxy, given the predicted and observed velocities.  The peculiar velocity is defined via \autoref{eqn:ModelRedshift}, such that a positive value indicates that the galaxy has a larger redshift than would be predicted from purely Hubble flow motion.}
\end{deluxetable*}

\section{Summary and discussion} \label{sec:Conclusions}

We have applied an improved approach for fitting maser data to obtain more precise distance estimates to four previously-published MCP galaxies: UGC 3789, NGC 6264, NGC 6323, and NGC 5765b.  We find that previous maser disk modeling efforts have typically overestimated the systematic measurement uncertainties associated with maser positions from VLBI maps.  By incorporating these error floors into our disk model as model parameters, we are able to fit for and then marginalize over them, eliminating a source of systematic uncertainty and generically improving the measurement precision.  We have combined the revised distance estimates for UGC 3789, NGC 6264, NGC 6323, and NGC 5765b with the recently-published distances to CGCG 074-064 and NGC 4258 -- both of which included error floors as parameters in the model -- to derive constraints on $H_0$.  Assuming a global velocity uncertainty of 250\,\kms associated with peculiar motions, we find $H_0 = 73.9 \pm 3.0$\,\kmsmpc.

Our fiducial $H_0$ measurement is determined exclusively using megamaser-based distance and velocity measurements, and it thus represents an independent cosmological probe from standard candles, gravitational lenses, and the CMB.  The primary source of systematic uncertainty in this measurement comes from the unknown peculiar motions of the maser galaxies, and we have considered three different treatments\footnote{We note that a variety of peculiar velocity correction schemes are possible beyond what we have explicitly tested in this paper, including compound schemes that combine two or more of the above methods.} for determining how these peculiar velocities could modify the $H_0$ value:

\begin{enumerate}
    \item We permit the global velocity uncertainty to be a free parameter that we fit alongside $H_0$.
    \item We replace each galaxy's recession velocity with the velocity of the group that galaxy is a member of.
    \item We replace each galaxy's recession velocity with the velocity predicted by a galaxy flow model evaluated at that galaxy's location and redshift.  We use the 2M++, CF3, and M2000 peculiar velocity models.
\end{enumerate}

\noindent The first two of the above treatments modify the best-fit $H_0$ value by less than 1\,\kmsmpc, though the measurement precision suffers when permitting the global velocity uncertainty to be a free parameter because of the reduced degrees of freedom.  Using recession velocities from either the 2M++ or CF3 galaxy flow models systematically reduces the best-fit $H_0$ value by 2.1\,\kmsmpc, and the measurement precision improves because of the smaller uncertainty (150\,\kms) associated with the catalog velocities.  The recession velocities from the M2000 model, on the other hand, result in a substantially increased best-fit $H_0$ value, which is larger by 3.0\,\kmsmpc than the fiducial measurement and by 5.1\,\kmsmpc than the measurement made from correcting for peculiar velocities using the other two flow models.

We have also performed a series of leave-one-out jackknife tests for each of the different peculiar velocity treatments.  We find that the removal of any single galaxy from the sample never modifies the best-fit $H_0$ value by more than $1{\sigma}$, indicating that our measurement is not being unduly influenced by a single outlying value.

We test the prior empirical claim that the local value of $H_0$ exceeds the early-Universe value \citep[e.g.,][]{Riess_2019,Wong_2019,Verde_2019} by calculating the probability of rejecting the null hypothesis that our measurement does not exceed Planck's, i.e.,

\begin{equation}
P(H_0 \leq H_{0,\text{Planck}}) = \int_{-\infty}^{\infty} \mathcal{P}(H) \left[ \int_{-\infty}^{H} \mathcal{P}(H_0') \text{d}H_0' \right] \text{d}H \label{eqn:PlanckComparison} ,
\end{equation}

\noindent where $\mathcal{P}(H_0)$ is our measured posterior distribution for $H_0$ and we treat the Planck measurement probability distribution $\mathcal{P}(H)$ as a Gaussian with mean and standard deviation given by the published measurement of $H_{0,\text{Planck}} = 67.4 \pm 0.5$\,\kmsmpc \citep{Planck_2018}.  The value of $P(H_0 \leq H_{0,\text{Planck}})$ is listed for all four peculiar velocity treatments in \autoref{tab:FitStatistics}.  The first treatment, our default, gives a 2\% chance that our value is lower than Planck's, corroborating the sense of the present tension in $H_0$ at 98\% confidence.  The other peculiar velocity treatments give confidences of 95--99\%.  Performing an analogous comparison with a late-Universe measurement from SH0ES of $H_{0,\text{SH0ES}} = 74.03 \pm 1.42$\,\kmsmpc \citep{Riess_2019},

\begin{equation}
P(H_0 \geq H_{0,\text{SH0ES}}) = \int_{-\infty}^{\infty} \mathcal{P}(H) \left[ \int_{H}^{\infty} \mathcal{P}(H_0') \text{d}H_0' \right] \text{d}H \label{eqn:SH0ESComparison} ,
\end{equation}

\noindent we find that our result is consistent with the SH0ES measurement, with little preference for a higher or lower value (see \autoref{tab:FitStatistics}).

The ${\sim}4\%$ $H_0$ constraint presented in this paper comes from consideration of only six megamaser-hosting galaxies, and the precision is ultimately limited by the quality and quantity of the available distance measurements.  Future $H_0$ measurements from the MCP will improve on this precision by incorporating distance measurements from additional megamaser-hosting galaxies.

\acknowledgments

We thank R.~B.~Tully and R.~Graziani for help with the \textit{Cosmicflows-3} peculiar velocities, M.~J.~Hudson for advice regarding the 2M++ peculiar velocities, L.~Blackburn for modeling discussions, and Erik Peterson for help with understanding redshift uncertainties.  The National Radio Astronomy Observatory is a facility of the National Science Foundation operated under cooperative agreement by Associated Universities, Inc.  This work made use of the Swinburne University of Technology software correlator, developed as part of the Australian Major National Research Facilities Programme and operated under license.  This research has made use of the NASA/IPAC Extragalactic Database (NED), which is operated by the Jet Propulsion Laboratory, California Institute of Technology, under contract with the National Aeronautics and Space Administration.  This work was supported in part by the Black Hole Initiative at Harvard University, which is funded by grants from the John Templeton Foundation and the Gordon and Betty Moore Foundation to Harvard University.

\vspace{5mm}
\facilities{GBT, VLA, VLBA, HSA}
\software{AIPS, CASA, GBTIDL, PyMC3 \citep{Salvatier2016}}, \texttt{dynesty} \citep{Speagle_2019}

\clearpage

\bibliography{mybib}{}
\bibliographystyle{aasjournal}

\appendix

\section{Updated disk modeling parameter values} \label{app:DiskModelingResults}

In \autoref{tab:DiskFittingResults} we list the values for all model parameters from the updated disk fits to UGC 3789, NGC 6264, NGC 6323, and NGC 5765b.  A comprehensive description of the model is given in \cite{Pesce_2020}.

\begin{deluxetable}{LcCCCC}
\tablecolumns{6}
\tablewidth{0pt}
\tablecaption{Updated disk fitting results\label{tab:DiskFittingResults}}
\tablehead{  & & \multicolumn{4}{c}{Galaxy} \\ \cline{3-6}
 \colhead{Parameter}	&	\colhead{Units}	&	\colhead{UGC 3789}	&	\colhead{NGC 6264} & \colhead{NGC 6323} & \colhead{NGC 5765b}}
\startdata
D &	Mpc & 51.5^{+4.5}_{-4.0} & 132.1^{+21.2}_{-17.3} & 109.4^{+34.2}_{-23.4} & 112.2^{+5.4}_{-5.1} \\
M_{\text{BH}} &	$10^7$ M$_{\odot}$ & 1.19^{+0.10}_{-0.09} & 2.76^{+0.45}_{-0.36} & 1.02^{+0.32}_{-0.22} & 4.15^{+0.20}_{-0.19} \\
v &	\kms & 3319.9 \pm 0.8 & 10192.6 \pm 0.8 & 7801.5 \pm 1.5 & 8525.7 \pm 0.7 \\
x_0 & mas & -0.4014 \pm 0.0010 & 0.0050 \pm 0.0012 & 0.0161 \pm 0.0010 & -0.0440 \pm 0.0014 \\
y_0 & mas & -0.4615 \pm 0.0011 & 0.0076 \pm 0.0016 & 0.0073 \pm 0.0024 & -0.0995 \pm 0.0019 \\
i_0 & degree & 84.9 \pm 0.6 & 91.3 \pm 2.3 & 91.5 \pm 0.3 & 72.4 \pm 0.5 \\
\frac{di}{dr} & degree\,mas$^{-1}$ & 7.7 \pm 1.0 & -1.5 \pm 4.0 & \ldots & 12.5 \pm 0.5 \\
\Omega_0 & degree & 222.4 \pm 0.4 & 84.7 \pm 1.3 & 184.4 \pm 0.6 & 149.7 \pm 0.3 \\
\frac{d\Omega}{dr} & degree\,mas$^{-1}$ & -1.6 \pm 0.6 & 16.7 \pm 2.2 & 12.9 \pm 1.2 & -3.2 \pm 0.2 \\
\midrule
\sigma_x & mas & 0.0045 \pm 0.0011 & 0.0012^{+0.0011}_{-0.0008} & 0.0035 \pm 0.0010 & 0.0028^{+0.0012}_{-0.0011} \\
\sigma_y & mas & 0.0063 \pm 0.0013 & 0.0056 \pm 0.0020 & 0.0037^{+0.0025}_{-0.0022} & 0.0034 \pm 0.0009 \\
\sigma_{v,\text{sys}} & \kms & 1.7^{+0.9}_{-0.8} & 1.6^{+0.9}_{-0.8} & 2.1^{+1.0}_{-0.9} & 1.1 \pm 0.5 \\
\sigma_{v,\text{hv}} & \kms & 1.8^{+0.9}_{-0.7} & 1.3^{+0.7}_{-0.6} & 1.9^{+0.9}_{-0.7} & 1.5^{+0.6}_{-0.5} \\
\sigma_a & \kmsyr & 0.34^{+0.06}_{-0.05} & 0.08^{+0.07}_{-0.05} & 0.21 \pm 0.09 & 0.041 \pm 0.014 \\
\enddata
\tablecomments{\textit{Top}: Fitting results for the global parameters describing the maser disk, marginalized over all other parameters; for each value we quote the posterior median and 1$\sigma$ confidence interval (i.e., 16th to 84th percentile).  Here, $D$ is the angular-size distance to the galaxy, $M_{\text{BH}}$ is the mass of the SMBH, $v$ is the line-of-sight CMB-frame velocity of the SMBH, $(x_0,y_0)$ is the coordinate location of the SMBH, $i_0$ is the inclination angle of the disk at $r=0$, $\frac{di}{dr}$ is the first-order inclination angle warping parameter, $\Omega_0$ is the position angle of the disk at $r=0$, and $\frac{d\Omega}{dr}$ is the first-order position angle warping parameter.  For the uncertainties we quote 1$\sigma$ (i.e., 16\% and 84\%) confidence intervals from the posteriors.  The SMBH coordinate locations are referenced to the coordinate zeropoint used in the respective data paper:  for UGC 3789, see \cite{Reid_2009}; for NGC 6264, see \cite{Kuo_2013}; for NGC 6323, see \cite{Kuo_2015}; and for NGC 5765b, see \cite{Gao_2017}.  \textit{Bottom}: Fitting results for the error floor parameters; $\sigma_x$ is the $x$-position error floor, $\sigma_y$ is the $y$-position error floor, $\sigma_{v,\text{sys}}$ is the error floor for the systemic feature velocities, $\sigma_{v,\text{hv}}$ is the error floor for the high-velocity feature velocities, and $\sigma_a$ is the acceleration error floor.}
\end{deluxetable}

\end{document}